\documentclass[aps,prl,twocolumn,showpacs,10pt,superscriptaddress,preprintnumbers,nofootinbib]{revtex4-1}
\usepackage{epsfig,amssymb,amsmath,psfrag,epstopdf,color}
\pdfoutput=1
\usepackage{graphicx}
\usepackage[margin=7pt,justification=centerlast]{caption}
\usepackage{subcaption}
 
\usepackage{paralist}
\usepackage{hyperref}
\allowdisplaybreaks

\def\beq{\begin{equation}}
\def\eeq{\end{equation}}
\def\bsp#1\esp{\begin{split}#1\end{split}}
\newcommand{\be}{\begin{equation}}
\newcommand{\ee}{\end{equation}}
\newcommand{\bea}{\begin{eqnarray}}
\newcommand{\eea}{\end{eqnarray}}

\def\Fig#1{Fig.~{\ref{#1}}}

\def\cO{{\mathcal O}}

\def\to{\rightarrow}

\def\ksl{\not{\hbox{\kern-2.3pt $k$}}}

\def\Ord{{\cal O}}

\def\spa#1.#2{\left\langle#1\,#2\right\rangle}
\def\spb#1.#2{\left[#1\,#2\right]}
\def\lor#1.#2{\left(#1\,#2\right)}
\def\sand#1.#2.#3{%
\left\langle\smash{#1}{\vphantom1}^{-}\right|{#2}%
\left|\smash{#3}{\vphantom1}^{-}\right\rangle}

\newcommand{\nn}{\nonumber}

\newcommand{\nbrk}{\nonumber\\}

\newcommand{\cI}{{\cal I}}

\begin{document}

\title{The Transverse Energy-Energy Correlator in the Back-to-Back Limit}
\author{AnJie Gao}
\affiliation{Zhejiang Institute of Modern Physics, Department of
  Physics, Zhejiang University, Hangzhou, 310027, China\vspace{0.5ex}}
\author{Hai~Tao~Li}
\affiliation{Theoretical Division, MS B283, Los Alamos National Laboratory, Los Alamos, NM 87545, USA\vspace{0.5ex}}
\author{Ian~Moult}
\affiliation{Berkeley Center for Theoretical Physics, University of California, Berkeley, CA 94720, USA\vspace{0.5ex}}
\affiliation{Theoretical Physics Group, Lawrence Berkeley National Laboratory, Berkeley, CA 94720, USA\vspace{0.5ex}}
\author{Hua~Xing~Zhu}
\affiliation{Zhejiang Institute of Modern Physics, Department of
  Physics, Zhejiang University, Hangzhou, 310027, China\vspace{0.5ex}}

\begin{abstract}
We present an operator based factorization formula for the transverse energy-energy correlator (TEEC) hadron collider event shape in the back-to-back (dijet) limit. This factorization formula exhibits a remarkably symmetric form, being a projection onto a scattering plane of a more standard transverse momentum dependent factorization. Soft radiation is incorporated through a dijet soft function, which can be elegantly obtained to next-to-next-to-leading order (NNLO) due to the symmetries of the problem. We present numerical results for the TEEC resummed to next-to-next-to-leading logarithm (NNLL) matched to fixed order at the LHC. Our results constitute the first NNLL resummation for a dijet event shape observable at a hadron collider, and the first analytic result for a hadron collider dijet soft function at NNLO. We anticipate that the theoretical simplicity of the TEEC observable will make it indispensable for precision studies of QCD at the LHC, and as a playground for theoretical studies of factorization and its violation. 
\end{abstract}

\maketitle

%%%%%%%%%%%%%%%%%%%%%%%%%%%%%%%%%%%%%%%%%%%%%%%%%%%%%%%%%%%%%%%%%%%%%%%%%%%%%%%%
\section{Introduction}
\label{sec:introduction}
%%%%%%%%%%%%%%%%%%%%%%%%%%%%%%%%%%%%%%%%%%%%%%%%%%%%%%%%%%%%%%%%%%%%%%%%%%%%%%%%

Event shape observables, which measure the flow of radiation in a scattering event, play a central role in QCD. They allow for precision measurements of QCD parameters, such as the strong coupling constant, $\alpha_s$, as well as for probes of more subtle features of QCD, such as color evolution or factorization violation.  While event shape observables in $e^+e^-$ collisions are by now quite well understood, with calculations incorporating next-to-next-to leading order (NNLO) fixed order corrections \cite{GehrmannDeRidder:2007hr,Gehrmann-DeRidder:2007nzq,Weinzierl:2008iv,Weinzierl:2009ms}, and next-to-next-to-next-to leading logarithmic (N$^3$LL) resummation \cite{Becher:2008cf,Abbate:2010xh,Chien:2010kc,Hoang:2014wka},  the same level of understanding has not been achieved for event shape observables at hadron colliders. This is due both to the technical complexity of fixed order calculations with multiple legs, and to the failure of standard factorization formulas in the hadron collider context. The theoretical and experimental study of event shape observables at hadron colliders therefore provides genuinely new opportunities for improving our understanding of QCD.

An important aspect in the description of event shapes is the resummation of singular terms in kinematic limits. For hadron collider event shapes, NNLL resummation has been achieved for zero-jet \cite{Stewart:2010pd,Becher:2015lmy,Becher:2015gsa} and one-jet event shapes \cite{Jouttenus:2013hs}. However, many interesting effects, namely non-trivial color evolution and amplitude level factorization violation, first occur for dijet event shapes, for which complete results are only available at NLL \cite{Kidonakis:1998bk,Kidonakis:1998nf,Banfi:2004nk,Banfi:2010xy,Sun:2014gfa,Hornig:2016ahz}.

A number of recent developments, namely the calculation of the three loop soft anomalous dimension \cite{Almelid:2015jia,Almelid:2017qju}, progress towards three jet production at NNLO \cite{Badger:2013gxa,Gehrmann:2015bfy,Dunbar:2016aux,Abreu:2017hqn,Badger:2017jhb,Badger:2018enw,Abreu:2018jgq,Abreu:2018zmy,Abreu:2018aqd,Chicherin:2018yne}, the illustration of the non-cancellation of Glauber effects in dijet processes~\cite{Collins:2007nk,Collins:2007jp,Mulders:2011zt}, the elucidation of amplitude-level factorization violation~\cite{Catani:2011st,Forshaw:2012bi,Schwartz:2017nmr},  and a formalism for the incorporation of factorization violation in the soft collinear effective theory (SCET) \cite{Rothstein:2016bsq}, motivate a renewed interest in the theoretical study of dijet event shapes.% \corr{In particular, we believe that there is good prospects of performing precision measurement, as well as exploring color evolution and factorization violation, in realistic dijet event shape observables at the LHC.}

%%%%%%%%%%%%%%%%%%%%%%%%%%%%%%%
\begin{figure}
\includegraphics[width=0.685\linewidth]{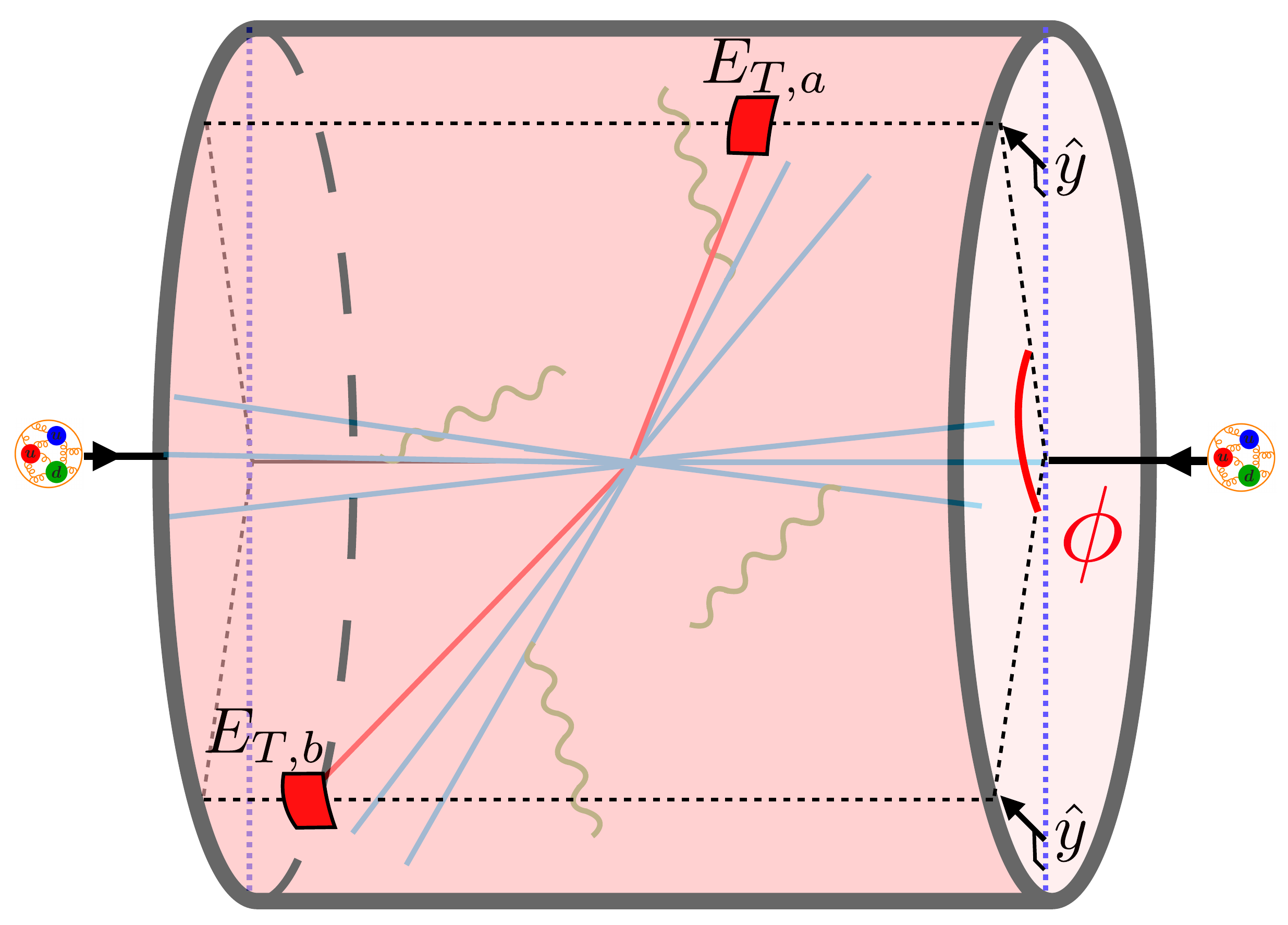}
\caption{The TEEC measures the $E_T$ weighted angular correlation of pairs of particles as a function of the angle $\phi$ in the transverse plane. In the $\phi \to \pi$ limit, it measures the momentum in the direction $\hat y$ perpendicular to the scattering plane spanned by the beam and jet axes, outlined in dashed blue.
}
\label{fig:TEEC}
\end{figure}
%%%%%%%%%%%%%%%%%%%%%%%%%%%%%%%

In this Letter,  we will study the transverse energy-energy-correlator (TEEC) observable \cite{Basham:1978bw,Ali:1984yp},
\begin{align}%\label{eq:TEEC_intro}
\text{TEEC}=\sum\limits_{a,b} \int d\sigma_{pp\to a+b+X} \frac{2 E_{T,a} E_{T,b}}{ |\sum_i E_{T,i}|^2 }   \delta(\cos\phi_{ab} - \cos\phi)\,, \nonumber
\end{align}
where the sum is over all pairs of hadrons, $E_{T}$ is the transverse energy of the hadrons, and $\phi_{ab}$ the azimuthal angle between the hadrons, as illustrated in \Fig{fig:TEEC}. For recent measurements of the TEEC for jets, see \cite{ATLAS:2015yaa,Aaboud:2017fml}.
Building on significant recent progress in the understanding of the energy-energy corelator (EEC) observable \cite{Hofman:2008ar,Belitsky:2013xxa,Belitsky:2013bja,Belitsky:2013ofa,Goncalves:2014ffa,Moult:2018jzp,Dixon:2018qgp}, we will show that the TEEC exhibits a remarkable perturbative simplicity in the dijet limit, allowing for significant progress to be made in the understanding of hadron collider event shapes.

%%%%%%%%%%%%%%%%%%%%%%%%%%%%%%%%%%%%%%%%%%%%%%%%%%%%%%%%%%%%%%%%%%%%%%%%%%%%%%%%
\section{Factorization Formula}
\label{sec:formalism}
%%%%%%%%%%%%%%%%%%%%%%%%%%%%%%%%%%%%%%%%%%%%%%%%%%%%%%%%%%%%%%%%%%%%%%%%%%%%%%%%

One of the main results of this Letter is an operator based factorization formula, derived in SCET \cite{Bauer:2000ew, Bauer:2000yr, Bauer:2001ct, Bauer:2001yt}, describing the singular behavior of the TEEC observable in the $\phi \to \pi$, or more conveniently, the $\tau \equiv \sin^2((\pi - \phi)/2)\to 0$ limit. In this limit, the singular behavior of the observable is described by a dijet configuration, with collinear radiation along the beam and jet axes, as well as low energy soft radiation. The $\tau\to 0$ limit defines a scattering plane spanned by the beam axis and the axis of the outgoing jets (more precisely the transverse thrust axis). Collinear splittings and soft emissions recoil the particles correlated by the TEEC observable slightly from this plane, see \Fig{fig:TEEC}.

The simplicity of the TEEC lies in a relation between the azimuthal angle $\phi$ and the momentum perpendicular to the scattering plane, which we will denote as the $y$ component, as illustrated in \Fig{fig:TEEC}. Consider two final state particles $k_3$ and $k_4$, whose transverse energy correlation is to be measured. In addition to the transverse momentum off the scattering plane due to final-state collinear splittings, they obtain transverse momentum from the recoil of the total soft momentum $k_{s,y}$, and from the momenta $k_{1,y}$ and $k_{2,y}$ of the incoming particles which enter the hard scattering. In the $\tau \to 0$ limit, we have the relation
\begin{align}
  \label{eq:1}
  \tau = \frac{\left(\frac{k_{3,y}}{\xi_3}+\frac{k_{4,y}}{\xi_4}+k_{1,y}+k_{2,y}-k_{s,y}\right)^2}{4 P_T^2}+\ldots \,, 
\end{align}
where $\xi_3$ and $\xi_4$ are the respective longitudinal momentum fractions of the two measured final state particles relative to the two leading jet momentum $p_3$ and $p_4$, and $P_T$ is the transverse momentum of $p_3$ and $p_4$ relative to the beam axis.

The relationship in Eq.~\ref{eq:1} allows us to derive a factorization formula for the TEEC in the dijet limit in terms of standard transverse momentum dependent (TMD) beam and jet functions
\begin{widetext}
  \begin{align}
\frac{d\sigma^{(0)}}{d\tau}
 =&\  \frac{p_T}{16 \pi s^2 (1 + \delta_{f_3 f_4}) \sqrt{\tau}}\sum\limits_{\text{channels}} \frac{1}{N_{\text{init}}}\int \frac{dy_3 dy_4 dp_T^2}{\xi_1\xi_2} \int_{-\infty}^{\infty}\frac{db}{2\pi}e^{-2ib\sqrt{\tau} p_T} \mathrm{tr}\big[\mathbf{H}^{f_1 f_2 \to f_3 f_4}(p_T,y^*,\mu) \mathbf{S}(b, y^*, \mu,\nu) \big]\nn \\
&\ \cdot   B_{f_1/N_1}(b,\,\xi_1,\,\mu,\,\nu)\,B_{f_2/N_2}(b,\,\xi_2,\,\mu,\,\nu) J_{f_3}\left(b,\mu,\nu\right)
  J_{f_4}\left(b,\mu,\nu\right). 
\label{eq:master}
\end{align}
\end{widetext}
Here the superscript $(0)$ indicates that this formula describes all contributions to the cross section that scale like $1/\tau$ modulo logarithms, up to potentially factorization violating terms which occur first at N$^4$LO, and will be discussed shortly. 
This factorization formula is a sum over different $2\to 2$ partonic scattering channels $f_1(p_1) f_2(p_2) \to f_3(p_3) f_4(p_4)$, where $N_{\text{init}}$ is the corresponding spin- and color-averaged factor for each channel, $\sqrt{s}$ is the center-of-mass energy,  $y_3$, $y_4$, and $p_T$ are the rapidity and transverse momentum of the two leading partonic jets at the lowest order in perturbation theory, and $\xi_1 = p_T(e^{y_3} + e^{y_4})/\sqrt{s}$ and $\xi_2 = p_T(e^{-y_3} + e^{-y_4})/\sqrt{s}$ are the born-level initial-state momentum fractions. The dependence on the scattering channel is incorporated through the hard function $\mathbf{H}^{f_1 f_2 \to f_3 f_4}(p_T,y^*,\mu)$, which depends on the $p_T$ and the single jet rapidity $y^* = (y_3 - y_4)/2$ in the partonic center-of-mass frame.  Each of the functions in Eq.~\ref{eq:master} depends on a virtuality renormalization scale $\mu$, and a rapidity renormalization scale $\nu$ \cite{Chiu:2011qc,Chiu:2012ir}. The associated renormalization group (RG) equations allow for the resummation of logarithms of $\tau$.

The soft and collinear dynamics in the dijet limit are described by beam functions, $B$, jet functions, $J$ and a soft function, $S$.  The beam functions and jet functions in Eq.~\eqref{eq:master} are \emph{identical} to the well-known TMD beam functions and EEC jet functions \cite{Moult:2018jzp} (which are in turn related to the TMD fragmentation functions \cite{Collins:2011zzd,Echevarria:2016scs,LuoTMD}).  Therefore, the TEEC in the dijet limit provides a probe into both beam and jet TMD dynamics that is interesting to a broad community. Since the TMD beam and jet functions are standard objects, we do not discuss them further, but collect all the anomalous dimensions and matching coefficients in the supplementary material.  The TEEC soft function is new and will be discussed shortly.

The factorization formula in Eq.~\ref{eq:master} is expected to be violated at N$^4$LO by Glauber gluons \cite{Collins:1988ig} which couple the different beam and jet functions. While the cancellation of Glauber gluons was shown for color singlet transverse moment distributions in the seminal works of \cite{Collins:1981uk,Collins:1981va,Collins:1981ta,Collins:1984kg,Collins:1985ue,Collins:1988ig,Collins:1989gx}, it is expected that factorization should not hold for a dijet event shape \cite{Collins:2007nk,Collins:2007jp,Bomhof:2007su,Rogers:2010dm,Buffing:2013dxa,Gaunt:2014ska,Zeng:2015iba,Catani:2011st,Schwartz:2017nmr,Forshaw:2008cq,Forshaw:2006fk,Martinez:2018ffw,Angeles-Martinez:2016dph,Forshaw:2012bi,Angeles-Martinez:2015rna,Schwartz:2018obd,Rothstein:2016bsq}. Glauber contributions can potentially be incorporated in our formalism using \cite{Rothstein:2016bsq}, and indeed one of our primary motivations is to understand such violations by identifying a dijet observable with the simplest perturbative structure. Apart from a brief comment on the anomalous dimension of the soft function at N$^3$LO, we leave the study of violations of this factorization formula to future work, and restrict ourselves to NNLL accuracy where Eq.~\ref{eq:master} holds. 

%%%%%%%%%%%%%%%%%%%%%%%%%%%%%%%%%%%%%%%%%%%%%%%%%%%%%%%%%%%%%%%%%%%%%%%%%%%%%%%%
\section{Soft Function}
\label{sec:soft_function}
%%%%%%%%%%%%%%%%%%%%%%%%%%%%%%%%%%%%%%%%%%%%%%%%%%%%%%%%%%%%%%%%%%%%%%%%%%%%%%%%

\begin{figure}
\includegraphics[width=0.6\linewidth]{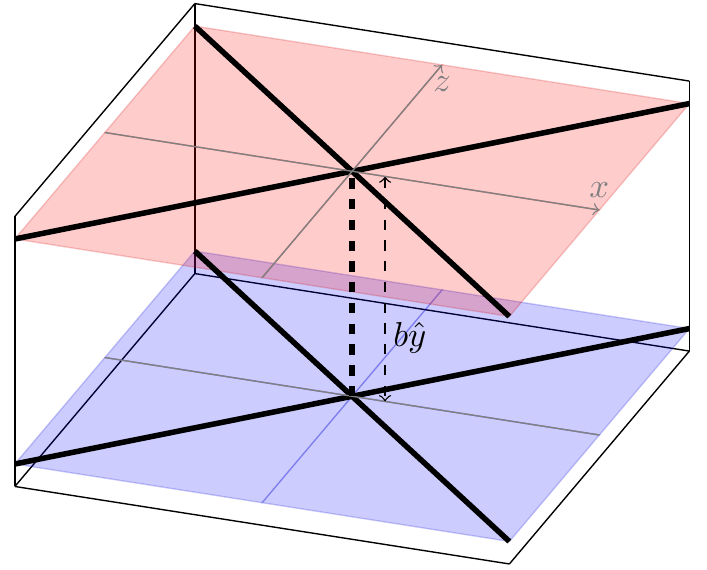}
\caption{The spatial structure of the TEEC soft function. Each set of Wilson lines lies in a scattering plane, and their relative displacement is perpendicular to these planes.
}
\label{fig:soft}
\end{figure}

The most complicated obstacle for precision calculations of multi-jet event shapes is the soft function, due to its dependence on multiple directions. (For recent progress towards numerical calculations of soft functions at NNLO, see \cite{Bell:2018vaa,Bell:2018oqa,Bell:2018mkk}.) A key feature of the TEEC which makes it particularly amenable to analytic higher order calculations is the simplicity of its soft function, which is defined as a vacuum expectation of Wilson lines,
\begin{align}
  \label{eq:soft}
\hspace{-0.25cm}\mathbf{S}(b,y^*) = \langle 0 |T[\boldsymbol{O}_{n_1 n_2 n_3 n_4}(0^\mu)] \overline{T}[\boldsymbol{O}_{n_1 n_2 n_3 n_4}^\dagger (b^\mu)] | 0 \rangle \,,
\end{align}
as illustrated in \Fig{fig:soft} (There the temporal direction has necessarily been suppressed). 
Here $\boldsymbol{O}_{n_1 n_2 n_3 n_4}(x) = \boldsymbol{Y}_{n_1} \boldsymbol{Y}_{n_2} \boldsymbol{Y}_{n_3} \boldsymbol{Y}_{n_4}(x)$, with $\boldsymbol{Y}_{n_i}(x) = \exp[ i \int ds\, n_i \cdot A(s n_i + x) \mathbf{T}_i]$ a semi-infinite light-like soft Wilson line, and $n_i^\mu = p_i^\mu/p_i^0$ the light-like direction of the incoming or outgoing parton in the partonic center-of-mass frame. The directions of the Wilson lines are standard and hence suppressed, as are gauge links at infinity. % In this frame, $n_1 \cdot n_2 = n_3 \cdot n_4 = 2$, and $n_1 \cdot n_3 = n_2 \cdot n_4 = 1 - \tanh y^*$. 
We have chosen coordinates such that $b^\mu = (0,0,b,0)$ is in the direction $\hat y$ perpendicular to the scattering plane, $\hat y \cdot n_i = 0$. 

The soft function defined in Eq.~\eqref{eq:soft} suffers from UV and rapidity divergences. Rapidity divergences are regulated using the exponential regulator of \cite{Li:2016axz}.  The soft function, which is a matrix in color space, satisfies the RG equation
\begin{align}
  \label{eq:softRG}
  \frac{d\mathbf{S}}{d\ln \mu^2} = \frac{1}{2} \left( \mathbf{\Gamma}_S^\dagger \cdot \mathbf{S} + \mathbf{S} \cdot \mathbf{\Gamma}_S \right) \,,
\end{align}
with \cite{Kidonakis:1998bk,Kidonakis:1998nf,Aybat:2006mz,Aybat:2006wq}
\begin{align}
  \label{eq:Gammas}
  \mathbf{\Gamma}_S = \sum_{i<j} \mathbf{T}_i \cdot \mathbf{T}_j \gamma_{\rm cusp} \ln\frac{\nu^2\, n_i \cdot n_j}{2 \mu^2} - \sum_i \frac{c_i}{2} \gamma_s \mathbf{1} - \boldsymbol{\gamma}_{\rm quad}  \,,
\end{align}
where $\nu$ is the rapidity scale, and $c_i = C_F$ or $C_A$ is the Casimir of the parton $i$. Here $\gamma_{\rm cusp}$ is the cusp anomalous dimension \cite{Korchemsky:1987wg}, $\gamma_s$ is the threshold soft anomalous dimension \cite{Li:2014afw} and  $\boldsymbol{\gamma}_{\rm quad}$ is the  anomalous dimension for quadrupole color and kinematic entanglement, which first appears at three loops \cite{Almelid:2015jia,Almelid:2017qju}. The evolution equation associated with the rapidity scale $\nu$ is
\begin{align}
  \label{eq:softnuRG}
  \frac{d \mathbf{S}}{d \ln \nu^2} =
\frac{1}{2} \left( \mathbf{\Gamma}_y^\dagger \cdot \mathbf{S} + \mathbf{S} \cdot \mathbf{\Gamma}_y \right) \,,
\end{align}
with
\begin{align}
\label{eq:nuAD}
\mathbf{\Gamma}_y = &\, \left(
\int_{\mu^2}^{b_0^2/b^2} \frac{d\bar{\mu}^2}{\bar{\mu}^2} \gamma_{\rm cusp} [\alpha_s(\bar{\mu})] + \gamma_r[\alpha_s(b_0/b)] \right) \sum_i c_i  \mathbf{1} \nn\\
&\, + \boldsymbol{\gamma}_X[y^*, \alpha_s(b_0/b)] \,.
\end{align}
This is the generalization of the rapidity RGE~\cite{Chiu:2011qc,Chiu:2012ir} for color singlet production to dijet production at hadron colliders. Here $\gamma_r$ is the rapidity anomalous dimension for the color transverse momentum distribution \cite{Li:2016ctv}, and $b_0=2e^{-\gamma_E}$.

%%%%%%%%%%%%%%%%%%%%%%%%%%%%%%%
%%%%%%%%%%%%%%%%%%%%%%%%%%%%%%%

The color non-diagonal rapidity anomalous dimension, $\boldsymbol{\gamma}_X$, vanishes at one and two-loops due to rescaling invariance, $n_i \to e^{\lambda_i} n_i$, which is sufficient for the NNLL resummation considered in this Letter. $\boldsymbol{\gamma}_X$ can potentially be non-zero at three loops where there is a scaling invariant cross ratio $n_1 \cdot n_3\, n_2 \cdot n_4 /(n_1 \cdot n_2\, n_3 \cdot n_4) = (1 - \tanh y^*)^2/4$.  The consistency of the factorization formula (derived from rapidity scale independence of the cross section) implies $\boldsymbol{\gamma}_X=0$ to all perturbative orders, however, since the factorization formula is expected to be violated, we do not take this as given. If $\boldsymbol{\gamma}_X=0$, it requires a symmetry explanation, and if not, it provides a direct window into factorization violation. Either way, we believe that the calculation of the TEEC soft function at three loops will provide considerable insight into rapidity factorization.

While the RG can be used to predict the logarithmic dependence of the soft function, its simple structure implies that the constants can also be easily computed. Writing its perturbative expansion as $\mathbf{S}=\sum(\alpha_s/4\pi)^n\mathbf{S}^{(n)}$, we have the beautiful relation
\begin{align}
  \mathbf{S}^{(1)}(y^*, L_b, L_\nu) =&  - \sum_{i<j} \left(\mathbf{T}_i \cdot \mathbf{T}_j \right) S_\perp^{(1)}\left(L_b, L_\nu + \ln \frac{n_i \cdot n_j}{2} \right) \,, \nonumber \\
\mathbf{S}^{(2)}(y^*, L_b, L_\nu) = &  - \sum_{i<j} \left(\mathbf{T}_i \cdot \mathbf{T}_j \right) S_\perp^{(2)}\left(L_b, L_\nu + \ln \frac{n_i \cdot n_j}{2} \right) \,, \nonumber \\
&+\frac{1}{2!} \left( \mathbf{S}^{(1)}(y^*, L_b, L_\nu)\right)^2 \,,
\end{align}
where $S_\perp^{(n)}(L_b, L_\nu)$ is the $n$-loop TMD soft function for color-singlet production at hadron colliders (which can be found up to three loops in \cite{Li:2016ctv}), and $L_b=\ln(\mu^2b^2/b_0^2)$, $L_\nu = \ln (\nu^2 b^2/b_0^2)$. This is the first analytic result for a hadron collider dijet soft function at NNLO (The 2-jettiness soft function was computed numerically in \cite{Bell:2018mkk}). The remarkable simplicity of the TEEC soft function should be compared with the soft functions for the $N$-jettiness observable \cite{Jouttenus:2011wh,Boughezal:2015eha,Campbell:2017hsw,Li:2018tsq,Bell:2018mkk}, which already at one-loop, can only be computed numerically. The reason for this simplicity is interesting, and deserves further comment. A soft function describes the expected value of radiation sourced by a configuration of Wilson lines, projected onto some direction(s). For the $N$-jettiness observable \cite{Stewart:2010tn}, these directions are the Wilson line directions themselves, which necessitates a partitioning of the phase space around the Wilson lines and leads to a complicated structure. For a dijet configuration, there is a unique direction perpendicular to the scattering plane defined by the four Wilson lines, which we have denoted $\hat y$, such that $\hat y\cdot n_i =0$ for all Wilson line directions $n_i$. This is the direction that is used to define the TEEC soft function, as shown in \Fig{fig:soft}, and leads to its simplicity. In particular, it implies that the scale independent part of the TEEC soft function can only be function of scaling invariant cross ratio of $n_i$. This points to the TEEC soft function as the uniquely simple dijet soft function, and we believe this simplicity will facilitate further analytic studies. 

%%%%%%%%%%%%%%%%%%%%%%%%%%%%%%%%%%%%%%%%%%%%%%%%%%%%%%%%%%%%%%%%%%%%%%%%%%%%%%%%
\section{Numerical Results}
\label{sec:results}
%%%%%%%%%%%%%%%%%%%%%%%%%%%%%%%%%%%%%%%%%%%%%%%%%%%%%%%%%%%%%%%%%%%%%%%%%%%%%%%%

We can use our factorization formula in Eq.~\eqref{eq:master} to present numerical results for the LHC at $\sqrt{s} = 13\,$TeV.  We use the anti-$k_T$ algorithm \cite{Cacciari:2008gp} with cone size $R =0.4$ to select events with two leading jets having averaged jet $P_T \geq 250\,$GeV and individual jet rapidity $|Y|<2.5$. The TEEC is computed for particles with rapidity $|y|<2.5$. Throughout, we will use the PDF4LHC15$\_$nnlo$\_$mc \cite{Butterworth:2015oua} parton distribution functions, and we take $\alpha_s(M_Z) = 0.118$.

\begin{figure}[h]
  \centering
  \includegraphics[width=\linewidth]{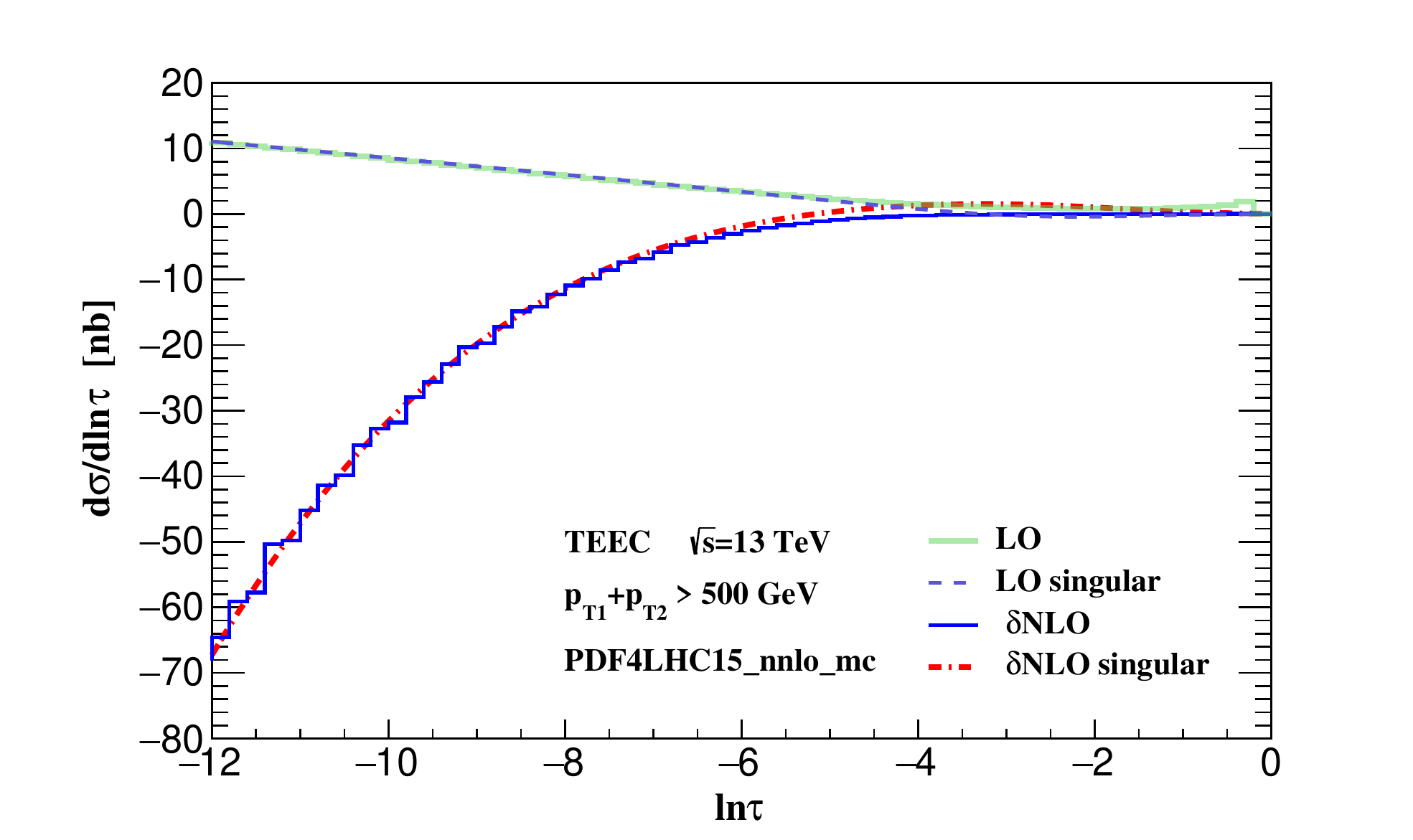}
  \caption{The TEEC at LO and NLO in the dijet limit. Here $\delta$NLO denotes only the NLO corrections.}
  \label{fig:asy}
\end{figure}

We begin by verifying that our factorization formula correctly reproduces the singular behavior as $\tau\to 0$ by comparing to the numerical code \textsc{Nlojet++} \cite{Nagy:2001fj,Nagy:2003tz}, which provides the LO and NLO QCD corrections to three-jet production. We note that NLO QCD corrections to the TEEC for jets have been computed in Ref.~\cite{Ali:2012rn} using \textsc{Nlojet++}, but here we are considering the TEEC for particles. Since the TEEC is first non-vanishing with a single emission from the dijet configuration, we use the perturbative counting for three-jet production for the matching. In Fig.~\ref{fig:asy} we show our factorization formula expanded to fixed order, compared with the numerical results of \textsc{Nlojet++} for $\tau d \sigma/d\tau$, finding perfect agreement. This is highly non-trivial, as both calculations are rather involved, with nine different partonic channels at LO, and provides a strong check on the validity of our factorization formula. To the best of our knowledge, this is the first time that the singular behavior for a dijet differential distribution is under full control at this order. 

\begin{figure}[h]
  \centering
  \includegraphics[width=\linewidth]{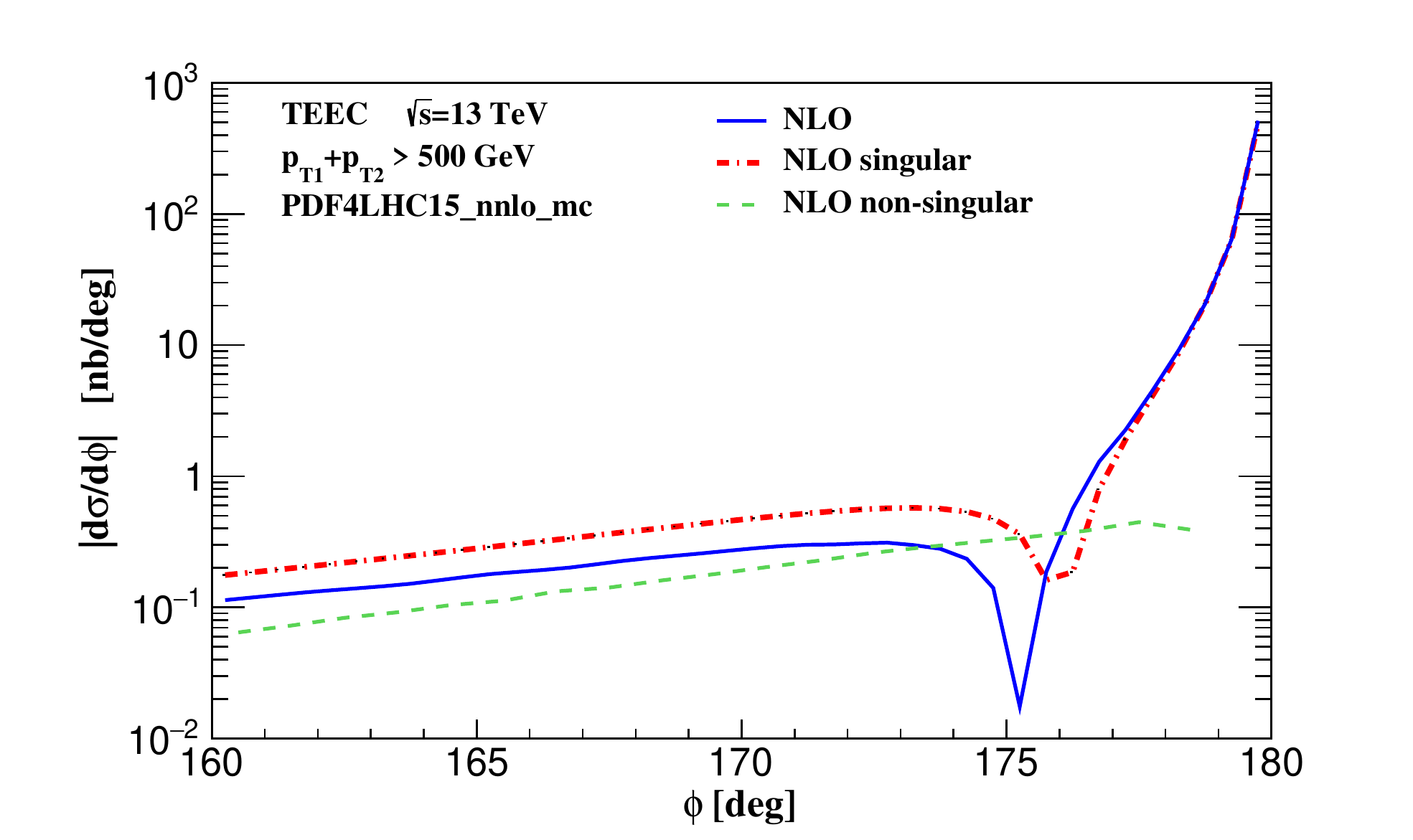}
  \caption{Fixed order singular and non-singular terms for the TEEC in the dijet limit.}
  \label{fig:powercorr}
\end{figure}

In Fig.~\ref{fig:powercorr}, we plot the full NLO prediction for $|d\sigma/d\phi|$ in the dijet limit, as well as its decomposition into the singular terms predicted by the factorization, and the non-singular terms (power corrections) defined as the difference between the full fixed order calculation and the singular result. For $\phi\to180^\circ$ the singular terms approach the full NLO predictions, as already demonstrated in Fig.~\ref{fig:asy}, but here we can more clearly see the interplay between the singular and non-singular terms. Since the TEEC effectively measures the $y$ component of an auxiliary transverse momentum $|q_y| \sim (\pi - \phi)$, this suggests that the power corrections start at $\Ord(\pi - \phi)$. It would be interesting to understand them further. Recent progress in the calculation of power corrections for transverse momentum type observables was made in \cite{Ebert:2018gsn}.

\begin{figure}%[h]
  \centering
  \includegraphics[width=\linewidth]{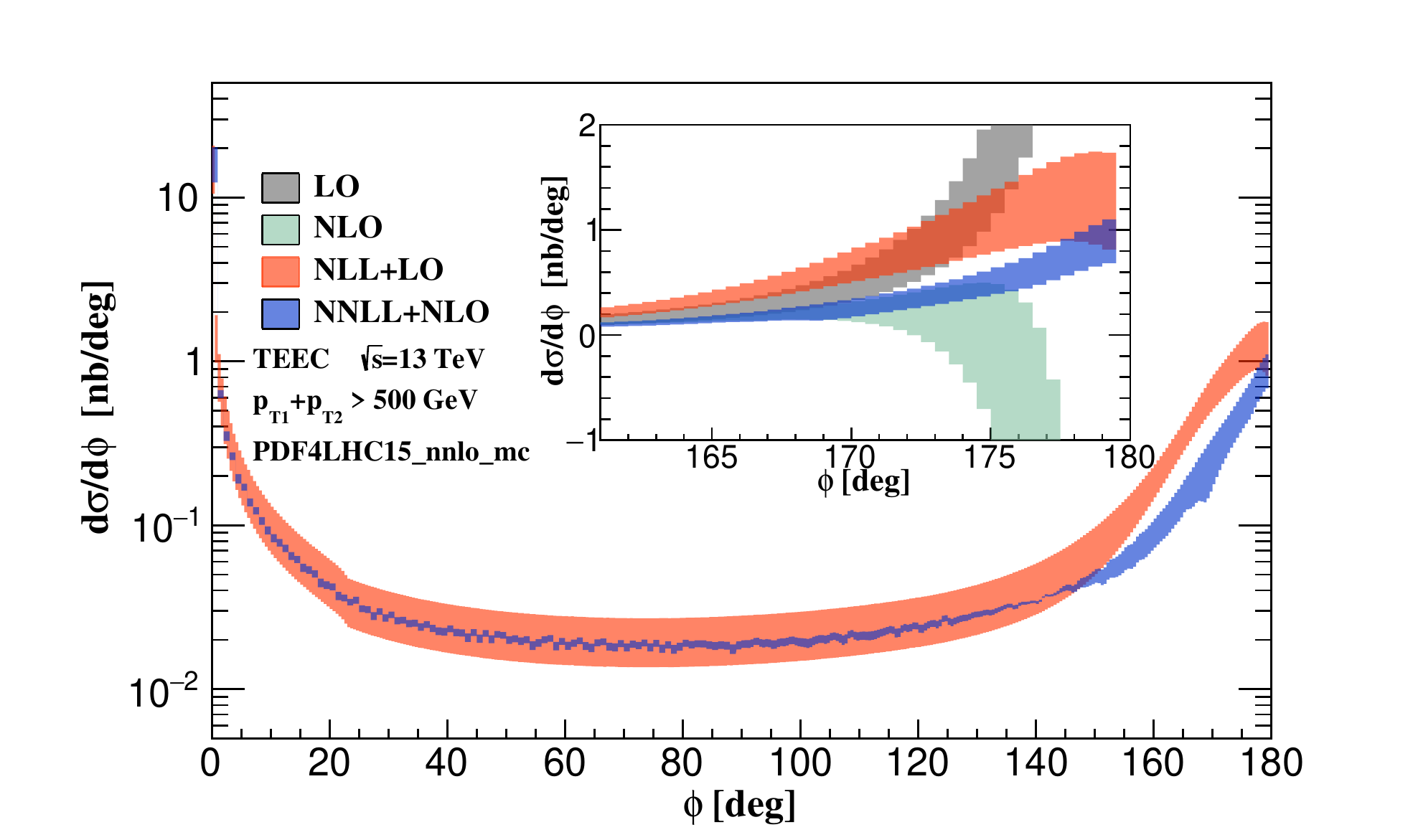}
  \caption{The resummed TEEC distribution matched to fixed order at both NLL+LO and NNLL+NLO.}
  \label{fig:resummed}
\end{figure}
In Fig.~\ref{fig:resummed} we show resummed predictions for the TEEC at NLL and NNLL, matched to LO and NLO, respectively. As can be seen from Fig.~\ref{fig:resummed}, the resummation cures the divergences in the fixed order calculations as $\phi\to 180^\circ$, and it would be particularly interesting to have precise experimental measurements in this region. Also clear is the reduction of scale uncertainties from NLL+LO to NNLL+NLO, although we find that the perturbative corrections are large. We leave a detailed analysis of various uncertainties coming from scale variation, matching, and non-perturbative corrections to future work.

%%%%%%%%%%%%%%%%%%%%%%%%%%%%%%%%%%%%%%%%%%%%%%%%%%%%%%%%%%%%%%%%%%%%%%%%%%%%%%%%
\section{Conclusions}
\label{sec:conclusion}
%%%%%%%%%%%%%%%%%%%%%%%%%%%%%%%%%%%%%%%%%%%%%%%%%%%%%%%%%%%%%%%%%%%%%%%%%%%%%%%%

In this Letter we have initiated the study of the TEEC hadron collider event shape. We have derived a factorization formula describing its singular behavior in the back-to-back (dijet) limit, and presented the first results for a dijet event shape at NNLL matched to NLO. The simplicity of the TEEC resides in its soft function, which we showed can be expressed in terms of a color singlet soft function through to NNLO.

There are a number of directions for further study and improvement. First, it will be interesting to compute the three-loop soft function for the TEEC to understand if $\boldsymbol{\gamma}_X$ is non-vanishing, and to understand the role of factorization violating terms at N$^3$LL. This will then enable matching to NNLO three-jet production once these become available \cite{Gehrmann:2015bfy,Dunbar:2016aux,Abreu:2017hqn,Badger:2017jhb,Badger:2018enw,Abreu:2018jgq,Abreu:2018zmy,Abreu:2018aqd,Chicherin:2018yne}. The resummation of collinear logarithms at $\phi\to 0$ can be performed systematically using an extension of the jet calculus \cite{Konishi:1979cb}, and will be described in a forthcoming work.  Finally, it would be interesting to compute the TEEC at strong coupling in planar $\mathcal{N}=4$ super Yang-Mills following \cite{Hofman:2008ar}, which could perhaps have relevance for heavy ion collisions. We believe the simplicity of the TEEC observable provides a laboratory for precision studies of QCD at the LHC, and for studying the structure of factorization and factorization violation for hadron collider event shapes.

%%%%%%%%%%%%%%%%%%%%%%%%%%%%%%%%%%%%%%%%%%%%%%%%%%%%%%%%%%%%%%%%%%%%%%%%%%%%%%%%
\section{Acknowledgements}
%%%%%%%%%%%%%%%%%%%%%%%%%%%%%%%%%%%%%%%%%%%%%%%%%%%%%%%%%%%%%%%%%%%%%%%%%%%%%%%%

We thank Ben Nachman and Wei Wang for useful discussions. I.M. and H.X.Z. would like to express a special thanks to the Mainz Institute for Theoretical Physics (MITP) for its hospitality and support. A.J.G. and H.X.Z. are supported in part by the One Hundred Talent Program of Zhejiang University. H.T.L. is supported by the Los Alamos National Laboratory LDRD program.
I.M. is supported by the Office of High Energy Physics of the U.S. DOE under Contract No. DE-AC02-05CH11231.

\bibliography{TEEC_bib.bib}{}
\bibliographystyle{apsrev4-1}

%======================================================================
%======================================================================
%======================================================================
%======================================================================
%======================================================================
%======================================================================
%======================================================================
\newpage

\onecolumngrid
%\newpage
\appendix

\section*{Supplemental material}

In this supplemental material, we collect the relevant anomalous dimension and matching coefficients used for the results presented in the main article. 

\subsection{Anomalous Dimensions}
\label{sec:anomalous-dimensions}

All our anomalous dimensions, generically denoted as $\gamma[\alpha_s, \ldots]$, where the dots represent potential dependence on kinematic variables, can be expanded in terms of $\alpha_s$,
\begin{align}
\gamma[\alpha_s,\ldots] = \sum_{n=0}^\infty \left(\frac{\alpha_s}{4 \pi} \right)^{n+1} \gamma_n[\ldots]   \,.
\end{align}
The QCD beta function $\beta[\alpha_s] = -2 \alpha_s \sum_{n=0} (\alpha_s/(4 \pi))^{n+1} \beta_n$ through to three loops are given by~\cite{Tarasov:1980au, Larin:1993tp}
\begin{align}
  \beta_0 = &\, \frac{11 C_A}{3}-\frac{2 n_f}{3} \,,
\nbrk
\beta_1 = &\, \frac{34 C_A^2}{3}-\frac{10 C_A n_f}{3}-2 C_F n_f \,,
\nbrk
\beta_2 = &\, \frac{2857 C_A^2}{54} + C_F^2 n_f - \frac{205 C_F C_A n_f}{18} - \frac{1415 C_A^2 n_f}{54} + \frac{11 C_F n_f^2}{9} + \frac{79 C_A n_f^2}{54} \,.
\end{align}
The cusp anomalous dimension through to three loops are~\cite{Korchemsky:1987wg, Moch:2004pa}
\begin{align}
\label{eq:c1}
  \gamma_{0}^{\rm cusp} = & \, 4 \,,
\nbrk
  \gamma_{1}^{\rm cusp} = & \, C_A  \left(\frac{268}{9}-8 \zeta_2\right)-\frac{40  n_f}{9} \,,
\nbrk
  \gamma_{2}^{\rm cusp} = & \, C_A^2  \left(-\frac{1072 \zeta_2}{9}+\frac{88
      \zeta_3}{3}+88 \zeta_4+\frac{490}{3}\right)
+C_A  n_f \left(\frac{160 \zeta_2}{9}-\frac{112
    \zeta_3}{3}-\frac{836}{27}\right)
%\brk
+ C_F n_f \left(32 \zeta_3-\frac{110}{3}\right)-\frac{16
                 n_f^2}{27} \,.
\nn
\end{align}
The quark and gluon anomalous dimensions through to two loops are \cite{Moch:2005id,Moch:2005tm,Idilbi:2005ni,Idilbi:2006dg,Becher:2006mr}
\begin{align}
  \gamma^{q}_0 = & \, -3 C_F \,,
\nbrk 
  \gamma^{q}_1 = & \, C_A C_F \left(-11 \zeta_2+26
    \zeta_3-\frac{961}{54}\right)
+C_F^2 \left(12 \zeta_2-24 \zeta_3-\frac{3}{2}\right)+C_F n_f \left(2 \zeta_2+\frac{65}{27}\right) \,,
\nbrk
  \gamma_0^{g} = &\, - \beta_0 \,,
\nbrk
  \gamma_1^{g} = &\, \,C_A^2 \left(\frac{11 \zeta_2}{3}+2 \zeta_3-\frac{692}{27}\right)+C_A n_f \left(\frac{128}{27}-\frac{2 \zeta_2}{3}\right)+2 C_F n_f \,.
\end{align}
The soft anomalous dimension through to two loops is
\begin{align}
    \gamma_{0}^s = & \, 0 \,,
\nbrk
  \gamma_{1}^s = & \, C_A \left(\frac{22 \zeta_2}{3}+28 \zeta_3-\frac{808}{27}\right)+ n_f \left(\frac{112}{27}-\frac{4 \zeta_2}{3}\right) \,.
\end{align}
The quadrupole correlation term $\boldsymbol{\gamma}_{\rm quad}[\{n_i\},\alpha_s]$ is only need for resummation beyond NNLL so we do not show it here. It can be found in \cite{Almelid:2015jia,Almelid:2017qju}.
The anomalous dimension for quark or gluon beam~($\gamma_B$) and jet~($\gamma_J$) function can then be obtained using the RG invariance condition,
\begin{align}
2 \gamma_q - C_F \gamma_s + 2 \gamma_{B,q} = 0 \,,\qquad
2 \gamma_g - C_A \gamma_s + 2 \gamma_{B,g} = 0 \,,
\end{align}
and $\gamma_{J,q(g)} = \gamma_{B,q(g)}$. 
The rapidity anomalous dimension through to two loops is given by
\begin{align}
\gamma_0^r =& \, \gamma_0^s \,,
\nbrk
\gamma_0^r = & \, \gamma_1^s - 2 \zeta_2 \beta_0  \,.
\end{align}
The relation between $\gamma_r$ and $\gamma_s$ was uncovered in \cite{Li:2016ctv}, and was shown to be the consequence of conformal symmetry of the special Wilson loop configuration in \cite{Vladimirov:2016dll,Vladimirov:2017ksc}. Again, the quadrupole rapidity anomalous dimension $\boldsymbol{\gamma}_X[y^*,\alpha_s]$ vanishes at one and two loops, and is not needed for NNLL resummation.

\subsection{Hard Functions}

The hard functions,  $\mathbf{H}^{f_1 f_2 \to f_3 f_4}$, are the infrared finite part of the $f_1 f_2 \to f_3 f_4$ squared amplitude (For a more precise definition, and detail discussion, see e.g. \cite{Moult:2015aoa}). They can be extracted from the known one-loop \cite{Kunszt:1993sd} and two-loop \cite{Anastasiou:2000kg, Anastasiou:2000ue,Glover:2001af, Bern:2002tk, Bern:2003ck, Glover:2003cm,Glover:2004si, Freitas:2004tk} amplitudes. The NLO hard functions for all partonic channels can be found in \cite{Kelley:2010fn,Moult:2015aoa}, and the NNLO hard functions can be found in the form of  \textsc{Mathematica} file in Ref.~\cite{Broggio:2014hoa}. We use the results in Ref.~\cite{Broggio:2014hoa} in our calculation. Notice that the color basis in these references are different.

The hard function is a matrix in color space. Given a color-space basis $|I\rangle$ for the two-to-two partonic amplitudes, it can be expressed as $(\mathbf{H})_{IJ} = \langle I | \mathcal{M} \rangle \langle \mathcal{M}^\dagger | J \rangle$, where $\mathcal{M}$ is the corresponding UV renormalized and appropriately IR subtracted two-to-two massless amplitudes. The hard function obeys the Renormalization Group~(RG) equation, 
\begin{align}
  \label{eq:hard}
  \frac{d \mathbf{H}}{d\ln\mu^2}  = \frac{1}{2} \left(\mathbf{\Gamma}_H \cdot \mathbf{H} + \mathbf{H} \cdot \mathbf{\Gamma}_H^\dagger \right) \,,
\end{align}
where the hard anomalous dimension $\mathbf{\Gamma}_H$ can be written as
\begin{align}
  \mathbf{\Gamma}_H = -\sum_{i < j}\mathbf{T}_i \cdot \mathbf{T}_j \gamma_{\rm cusp}\ln \frac{\sigma_{ij} \hat{s}_{ij} + i0}{\mu^2} + \sum_i \gamma_i \mathbf{1}  + \boldsymbol{\gamma}_{\rm quad} \,, \nn
\end{align}
where $\mathbf{T}_i$ is color-insertion operator, $\sigma_{ij} = -1$ if both $i$ and $j$ are incoming or outgoing, and $1$ otherwise. $\hat{s}_{ij} = 2 p_i \cdot p_j$ is the Mandelstam variables. Here $\gamma_i = \gamma_q\,, \gamma_g$ are the quark or gluon anomalous dimension.

\subsection{Beam Functions}

Both the beam function and jet function satisfy the following RG and rapidity RG equations, 
\begin{align}
  \label{eq:jetRG}
  \frac{d  G_i}{d\ln \mu^2} = \left( - \frac{1}{2} c_i \gamma_{\rm cusp} \ln \frac{4 (p_i^0)^2}{\nu^2} + \gamma_{G,i} \right) G_i \,,
\end{align}
\begin{align}
  \label{eq:jetnuRG}
  \frac{d  G_i}{d\ln \nu^2} = \frac{c_i}{2}\left(\int_{b_0^2/b^2}^{\mu^2} \frac{d \bar{\mu}^2}{\bar{\mu}^2}  \gamma_{\rm cusp}[\alpha_s(\bar{\mu})] - \gamma_r[\alpha_s(b_0/b)] \right) G_i \,,
\end{align}
where $G$ stands for $B$ or $J$.

The TMD beam functions for the TEEC can be matched onto standard PDFs at small but perturbative transverse momentum, 
\begin{align}
B_{i/N}(b,\xi,\mu,\nu) =\sum \limits_j \int \frac{dz}{z}\cI_{ij}\left(z,L_b,L_Q\right) f_{j/N} \left(\frac{\xi}{z}, \mu\right)   + \text{power corrections} \,,
\end{align}
where $L_b = \ln (b^2 \mu^2/b^2_0)$, $b_0 = 2 e^{-\gamma_E}$, and $L_Q = \ln (Q^2/\nu^2)$, with $Q = 2 p_i^0$, twice the energy of the measured parton energy. Note that unlike the conventional TMDPDF, here the gluon TMD beam function  has only one tensor structure, which we choose to be $1$. The reason is that the beam function here measures transverse momentum only in the $x$ direction. The matching coefficients have been derived to two loops in \cite{Gehrmann:2012ze,Gehrmann:2014yya,Echevarria:2016scs,Luebbert:2016itl,LuoTMD}. All the TMD beam functions through one loop can be written as
\begin{align}
  \label{eq:15}
 \mathcal{I}_{qq}(z,L_b,L_Q) = \, & \delta(1-z) + \left(\frac{\alpha_s}{4 \pi}\right) \Big[ C_F \left( -2 L_b L_Q + 3 L_b\right)
                       \delta(1-z) - P_{0,\,qq}(z)L_b  + 2C_F
                       (1-z) \Big] + \cO(\alpha_s^2)\,,\nbrk
\mathcal{I}_{qg}(z,L_b,L_Q) = \, & \left(\frac{\alpha_s}{4 \pi}\right) \Big[ 2 z (  1 - z ) -P_{0,qg}(z) L_b \Big] + \cO(\alpha_s^2)\,,\nn\\
\mathcal{I}_{gq}(z,L_b,L_Q) =\,& \left(\frac{\alpha_s}{4 \pi}\right) \Big[ - P_{0,\,gq}(z)L_b  + 2C_F\, z  \Big] + \cO(\alpha_s^2) \,,\nn\\
\mathcal{I}_{gg}(z,L_b,L_Q) =\, & \delta(1-z) + \left(\frac{\alpha_s}{4 \pi}\right) \Big[ \left( -2C_A L_b L_Q + \beta_0 L_b\right)
                       \delta(1-z) - P_{0,\,gg}(z)L_b \Big] + \cO(\alpha_s^2)  \,.
\end{align}
where $P_{0,ij}(z)$
are the usual LO splitting functions
\begin{align}
  \label{eq:17}
  P_{0,qq}(z) =\,& C_F \left[ 3  \delta(1-z) + \frac{4}{\left[1-z\right]}_+
                    -2  (1+z) \right]\,,\nbrk
  P_{0,qg}(z) = \, & 1 - 2 z + 2 z^2\,,\nbrk
  P_{0,gq}(z) = \, & 2 C_F \left[\frac{1+(1-z)^2}{z}\right]\,,\nbrk
  P_{0,gg}(z) = \, & 4 C_A \left[\frac{z}{\left[1-z\right]}_++\frac{1-z}{z}+z(1-z)\right]+\beta_0\delta(1-z)\,.
\end{align}

\subsection{Jet Functions}
The TEEC jet functions are the same as for the  EEC \cite{Moult:2018jzp}
\begin{align}
  J_q(b,\mu,\nu)=\, & J_{\bar{q}}(b,\mu,\nu)=  1+\left(\frac{\alpha_s}{4\pi}\right) C_F (-2L_bL_Q+3L_b+4-8\zeta_2)  + \cO(\alpha_s^2) \,, \nbrk
  J_g(b,\mu,\nu)=  \, & 1+\left(\frac{\alpha_s}{4\pi}\right)\left[-2C_A L_b L_Q+\beta_0 L_b+\left(\frac{65}{18}-8\zeta_2\right)C_A-\frac{5}{18} n_f\right]  + \cO(\alpha_s^2) \,.
\end{align}

\subsection{Soft Function}

The TEEC soft function is a matrix in color space. Writing its perturbative expansion as
\begin{align}
\mathbf{S}(b,y^*, \mu,\nu) = \mathbf{1} + \frac{\alpha_s}{4 \pi} \mathbf{S}^{(1)}(y^*, L_b, L_\nu) + 
\left(\frac{\alpha_s}{4 \pi} \right)^2 \mathbf{S}^{(2)}(y^*, L_b, L_\nu) + \cO(\alpha_s^3) \,.
\end{align}
The one-loop coefficient is
\begin{align}
  \mathbf{S}^{(1)}(y^*, L_b, L_\nu) =  - \sum_{i<j} \left(\mathbf{T}_i \cdot \mathbf{T}_j \right) S_\perp^{(1)}\left(L_b, L_\nu + \ln \frac{n_i \cdot n_j}{2} \right) \,,
\end{align}
where $S_\perp^{(1)}(L_b, L_\nu)$ is the one-loop TMD soft function for color-singlet production at hadron collider,
\begin{align}
S_\perp^{(1)}(L_b, L_\nu) = 2 L_b^2 - 4 L_b L_\nu -2 \zeta_2 \,.  
\end{align}
The two-loop results for the TEEC soft function can also be easily determined to be
\begin{align}
\label{eq:soft2}
\mathbf{S}^{(2)}(y^*, L_b, L_\nu) = \frac{1}{2!} \left( \mathbf{S}^{(1)}(y^*, L_b, L_\nu)\right)^2   - \sum_{i<j} \left(\mathbf{T}_i \cdot \mathbf{T}_j \right) S_\perp^{(2)}\left(L_b, L_\nu + \ln \frac{n_i \cdot n_j}{2} \right) \,,
\end{align}
where the first term in Eq.~\eqref{eq:soft2} is due to Non-Abelian Exponentiation theorem \cite{Gatheral:1983cz,Frenkel:1984pz}, while the second term is the genuine two-loop correction, which, as explained in the text, can be expressed in terms of the two-loop TMD soft function, $S_\perp^{(2)}$~\cite{Li:2016ctv}.
%%%%%%%%%%%%%%%%%%%%%%%%

\end{document}